\documentclass[12pt]{article}
  \usepackage{amsfonts}
  \usepackage{amsmath}
\usepackage{amssymb}
\usepackage{amscd}
\usepackage[dvips]{graphicx}
\begin{document}
~~
\bigskip
\bigskip
\begin{center}
{\Large {\bf{{{Spectra of disc  operator  for twisted
acceleration-enlarged Newton-Hooke space-times}}}}}
\end{center}
\bigskip
\bigskip
\bigskip
\begin{center}
{{\large ${\rm {Marcin\;Daszkiewicz}}$ }}
\end{center}
\bigskip
\begin{center}
{ {{{Institute of Theoretical Physics\\ University of Wroc{\l}aw pl.
Maxa Borna 9, 50-206 Wroc{\l}aw, Poland\\ e-mail:
marcin@ift.uni.wroc.pl}}}}
\end{center}
\bigskip
\bigskip
\bigskip
\bigskip
\bigskip
\bigskip
\bigskip
\bigskip
\begin{abstract}
The time-dependent spectra of disc  area operator for twisted
acceleration-enlarged Newton-Hooke
 space-times are derived. It is demonstrated
that the corresponding area quanta are expanding or  oscillating in
time.
\end{abstract}
\bigskip
\bigskip
\bigskip
\bigskip
\bigskip
\bigskip
\bigskip
\bigskip
\bigskip
 \eject

Recently, there appeared a lot of papers dealing with noncommutative
classical and quantum  mechanics (see e.g. \cite{mech}-\cite{qmnext})
as well as with field theoretical models (see e.g. \cite{field},
\cite{fieldnext}), in
which  the quantum space-time is employed. The suggestion to use
noncommutative coordinates goes back to Heisenberg and was firstly
formalized by Snyder in \cite{snyder}. Recently, there were also
found formal arguments based mainly on Quantum Gravity \cite{grav1} 
and String Theory models \cite{string1}, 
indicating that space-time at Planck scale  should be
noncommutative, i.e. it should  have a quantum nature. On the other
side, the main reason for such considerations follows from the
suggestion  that relativistic space-time symmetries should be
modified (deformed) at Planck scale, while  the classical Poincare
invariance still remains valid at
larger distances \cite{1a}, \cite{1anext}.

Presently, it is well known, that in accordance with the
Hopf-algebraic classification of all deformations of relativistic
and nonrelativistic symmetries, one can distinguish three basic
types of quantum spaces \cite{class1}, \cite{class2}:\\
\\
{ \bf 1)} Canonical ($\theta^{\mu\nu}$-deformed) space-time
\begin{equation}
[\;{ x}_{\mu},{ x}_{\nu}\;] = i\theta_{\mu\nu}\;\;\;;\;\;\;
\theta_{\mu\nu} = {\rm const}\;, \label{noncomm}
\end{equation}
introduced in  \cite{oeckl}, \cite{chi}
in the case of Poincare quantum group and in \cite{dasz1} 
for its Galilean counterpart.\\
\\
{ \bf 2)} Lie-algebraic modification of classical space
\begin{equation}
[\;{ x}_{\mu},{ x}_{\nu}\;] = i\theta_{\mu\nu}^{\rho}{ x}_{\rho}\;,
\label{noncomm1}
\end{equation}
with  particularly chosen coefficients $\theta_{\mu\nu}^{\rho}$
being constants. This type of noncommutativity has been obtained as
the representations of the $\kappa$-Poincare \cite{kappaP} and
$\kappa$-Galilei
\cite{kappaG} as well as the  twisted relativistic \cite{lie1} 
and  nonrelativistic \cite{dasz1} 
symmetries, respectively. \\
\\
{ \bf 3)} Quadratic deformation of Minkowski and Galilei  space
\begin{equation}
[\;{ x}_{\mu},{ x}_{\nu}\;] = i\theta_{\mu\nu}^{\rho\tau}{
x}_{\rho}{ x}_{\tau}\;, \label{noncomm2}
\end{equation}
with coefficients $\theta_{\mu\nu}^{\rho\tau}$ being constants. This
kind of deformation has been proposed in \cite{qdef}, \cite{paolo},
\cite{lie1}
 at relativistic and in \cite{dasz1} at  nonrelativistic level.\\
\\
Besides, it has been demonstrated in \cite{nh}, that in the case of
so-called acceleration-enlarged Newton-Hooke Hopf algebras
$\,{\mathcal U}_0(\widehat{ NH}_{\pm})$ the (twist) deformation
provides the new  space-time noncommutativity, which is expanding
($\,{\mathcal U}_0(\widehat{ NH}_{+})$) or periodic ($\,{\mathcal
U}_0(\widehat{ NH}_{-})$) in time, i.e. it takes the
form\footnote{The $\,{\mathcal U}_0(\widehat{ NH}_{\pm})$
acceleration-enlarged Newton-Hooke Hopf structures are obtained by
adding to the ${\widehat{NH}}_{\pm}$ algebras (algebraic parts) the
trivial coproduct $\Delta_{0}(a) = a\otimes 1+ 1\otimes
a$.},\footnote{$x_0 = ct$.}
\begin{equation}
{ \bf 4)}\;\;\;\;\;\;\;\;\;[\;t,{ x}_{i}\;] = 0\;\;\;,\;\;\; [\;{ x}_{i},{ x}_{j}\;] = 
if_{\pm}\left(\frac{t}{\tau}\right)\theta_{ij}(x)
\;, \label{nhspace}
\end{equation}
with time-dependent  functions
$$f_+\left(\frac{t}{\tau}\right) =
f\left(\sinh\left(\frac{t}{\tau}\right),\cosh\left(\frac{t}{\tau}\right)\right)\;\;\;,\;\;\;
f_-\left(\frac{t}{\tau}\right) =
f\left(\sin\left(\frac{t}{\tau}\right),\cos\left(\frac{t}{\tau}\right)\right)\;,$$
and $\theta_{ij}(x) \sim \theta_{ij} = {\rm const}$ or
$\theta_{ij}(x) \sim \theta_{ij}^{k}x_k$. Such a kind  of
noncommutativity  follows from the presence in acceleration-enlarged
Newton-Hooke symmetries $\,{\mathcal U}_0(\widehat{ NH}_{\pm})$ of
the time scale parameter (cosmological constant) $\tau$. As it was
demonstrated in \cite{nh} that  just this parameter   is responsible
for oscillation or expansion of space-time noncommutativity.

It should be noted that both Hopf structures $\,{\mathcal
U}_0(\widehat{ NH}_{\pm})$ contain, apart from rotation $(M_{ij})$,
boost $(K_{i})$ and space-time translation $(P_{i}, H)$ generators,
the additional ones denoted by $F_{i}$, responsible for constant
acceleration. Consequently, if all generators
 $F_{i}$ are equal zero we obtain the twisted Newton-Hooke quantum
 space-times \cite{nh1}, while for time parameter $\tau$ running to infinity
we get the acceleration-enlarged twisted Galilei Hopf structures proposed in
\cite{nh}. Besides, for $F_i\to 0$ and $\tau \to \infty$
 we reproduce the canonical (\ref{noncomm}),
Lie-algebraic (\ref{noncomm1}) and quadratic (\ref{noncomm2})
(twisted) Galilei spaces provided in \cite{dasz1}.\\
Finally, it should be noted, that all  mentioned above
noncommutative space-times have been  defined as the quantum
representation spaces, so-called Hopf modules (see
\cite{bloch}, \cite{wess}, \cite{oeckl}, \cite{chi}), for quantum acceleration-enlarged
Newton-Hooke Hopf algebras, respectively.

 As it was already mentioned, the impact of quantum
spaces on the formal structure of  physical systems has been
discussed in different context in \cite{mech}-\cite{fieldnext}.
Particulary, it has been demonstrated that the different kinds of
noncommutativity produce additional interaction terms in the
Hamiltonian of dynamical systems, in particular, it has been shown,
that such quantum spaces generate  nonlocality. Besides, there were
also performed the studies on the fractal structure of quite popular
  $\kappa$-Minkowski space-time as well as the quantum sphere \cite{fractal}. More precisely,
  there has been shown that both spaces have a scale dependent
  fractal dimension,  which deviates from its classical value at short
  scales. It can be added, that the formal (algebraic) properties of $\kappa$-Poincare
  algebra gave also  a mathematical tool for
 such theoretical constructions as Double Special
Relativity (see e.g. \cite{dsr}), which postulates two
observer-independent scales, of velocity, describing the speed of
light, and of mass, which can be identify with $\kappa$-parameter -
the fundamental Planck mass.

Recently, an  interesting property of (non-)relativistic
(three-dimensional) canonically deformed space-time (\ref{noncomm})
\begin{equation}
[\;t,{ x}_{1,2}\;] = 0\;\;\;,\;\;\; [\;{ x}_{1},{ x}_{2}\;] = 
i\theta
\;, \label{spacesss}
\end{equation}
has been studied  in the article \cite{area}. It has been
demonstrated that the spectra of disc area operator
\begin{equation}
S \equiv \pi R^2 =  \pi \left( [\;x_1\;]^2 + [\;x_2\;]^2   \right)
\;, \label{thetaarea}
\end{equation}
takes the discrete values
\begin{equation}
s_n = 2\pi \theta ( n+{1}/{2} )\;\;;\;\; n =0, 1, 2, \ldots \;,
\label{canspec}
\end{equation}
i.e. it has been shown that there appear, in such a case,
proportional to the deformation parameter $\theta$,  quanta of area
$s_{\theta} = 2\pi {\theta}$. This result appears to be quite
interesting due to similarity to Quantum Gravity  area operator
\cite{QG}. In other words, such a result provides the link between
canonical space \cite{oeckl}, \cite{chi},  and the quantum gravitational
considerations performed in the framework of spin foam model
\cite{spinfoam} (more preciously, it provides the similar to
(\ref{canspec}) area spectrum \cite{spinfoam1}
\begin{equation}
s(j_k) = \gamma l_p^2 ( j_k+{1}/{2}) \;\;;\;\; j_k =0, 1/2, 1, 3/2,
2, \ldots \;, \label{spinspec}
\end{equation}
with Planck length $l_p = (\hbar G/c)^{1/2} = 10^{-33}$ cm and
constant  $\gamma$ analogous to the Immirzi parameter). \\
Another interesting result   has been obtained in \cite{snyderspec}
where the Bekenstein spectrum of black hole horizon area
\cite{bekenstein}
\begin{equation}
s_n = \tilde{\gamma} l_p^2 n \;\;;\;\; n =0, 1, 2, \ldots \;\;;\;\;
\tilde{\gamma} - \rm{dimensionless\;constant} \;, \label{bekspec}
\end{equation}
has been rederived as a spectrum of disc  area operator in quantum
Snyder space \cite{snyder}.

In this article we find, following the algorithm used in
\cite{area}, \cite{snyderspec}, the disc spectra of  area operator
 for all (with classical time and quantum space directions) three-dimensional acceleration-enlarged
Newton-Hooke quantum space-times provided in \cite{nh}. They can
be written as follows
\begin{equation}
[\;t,{ x}_{1,2}\;] = 0\;\;\;,\;\;\; [\;{ x}_{1},{ x}_{2}\;] = 
if({t})
\;, \label{spaces}
\end{equation}
where function $f({t})$ is given by\footnote{See formulas
(\ref{noncomm1}) and (\ref{nhspace}) respectively.}
\begin{eqnarray}
f({t})&=&f_{\kappa_1}({t}) =
f_{\pm,\kappa_1}\left(\frac{t}{\tau}\right) = \kappa_1\,C_{\pm}^2
\left(\frac{t}{\tau}\right)\;, \label{w2}\\
f({t})&=&f_{\kappa_2}({t}) =
f_{\pm,\kappa_2}\left(\frac{t}{\tau}\right) =\kappa_2\tau\, C_{\pm}
\left(\frac{t}{\tau}\right)S_{\pm} \left(\frac{t}{\tau}\right) \;,
\label{w3}\\
f({t})&=&f_{\kappa_3}({t}) =
f_{\pm,\kappa_3}\left(\frac{t}{\tau}\right) =\kappa_3\tau^2\,
S_{\pm}^2 \left(\frac{t}{\tau}\right) \;, \label{w4}\\
f({t})&=&f_{\kappa_4}({t}) =
 f_{\pm,\kappa_4}\left(\frac{t}{\tau}\right) = 4\kappa_4
 \tau^4\left(C_{\pm}\left(\frac{t}{\tau}\right)
-1\right)^2 \;, \label{w5}\\
f({t})&=&f_{\kappa_5}({t}) =
f_{\pm,\kappa_5}\left(\frac{t}{\tau}\right) = \pm \kappa_5\tau^2
\left(C_{\pm}\left(\frac{t}{\tau}\right)
-1\right)C_{\pm} \left(\frac{t}{\tau}\right)\;, \label{w6}\\
f({t})&=&f_{\kappa_6}({t}) =
f_{\pm,\kappa_6}\left(\frac{t}{\tau}\right) = \pm \kappa_6\tau^3
\left(C_{\pm}\left(\frac{t}{\tau}\right) -1\right)S_{\pm}
\left(\frac{t}{\tau}\right)\;, \label{w7}
\end{eqnarray}
for six types of acceleration-enlarged Newton-Hooke Hopf algebras
$\,{\mathcal U}_{\kappa_1}(\widehat{ NH}_{\pm})$, $\,{\mathcal
U}_{\kappa_2}(\widehat{ NH}_{\pm})$, $\,{\mathcal
U}_{\kappa_3}(\widehat{ NH}_{\pm})$, $\,{\mathcal
U}_{\kappa_4}(\widehat{ NH}_{\pm})$, $\,{\mathcal
U}_{\kappa_5}(\widehat{ NH}_{\pm})$ and $\,{\mathcal
U}_{\kappa_6}(\widehat{ NH}_{\pm})$ respectively, with
$$C_{+/-} \left(\frac{t}{\tau}\right) = \cosh/\cos \left(\frac{t}{\tau}\right)\;\;\;{\rm and}\;\;\;
S_{+/-} \left(\frac{t}{\tau}\right) = \sinh/\sin
\left(\frac{t}{\tau}\right) \;.$$
 Let us note that all above
nonrelativistic space-times are equipped with the  classical time
and quantum spatial directions. Of course, as it was already
mentioned above, for generators $F_i$ approaching  zero (and)
parameter $\tau$ running to infinity the space-times
(\ref{w2})-(\ref{w7}) become the same as twisted Newton-Hooke (and) Galilei
spaces proposed in \cite{nh1} (and) \cite{dasz1} respectively. In the
case $\tau \to \infty$ we get the acceleration-enlarged Galilei
quantum space-times proposed in \cite{nh}.

In a first step of our construction we define two time-dependent
operators $(i=1,\ldots,6)$
\begin{equation}
a_{\kappa_i}(t) = \frac{1}{\sqrt{2f_{\kappa_i}({t})}}\left(x_1 +
ix_2 \right)\;,\label{operators1}
\end{equation}
\begin{equation}
 a^{\dag}_{\kappa_i}(t) = \frac{1}{\sqrt{2f_{\kappa_i}({t})}}\left(x_1 - ix_2
\right)  \;. \label{operators2}
\end{equation}
One can check that they satisfy the following (standard) commutation
relations
\begin{equation}
[\;a_{\kappa_i}(t),a^{\dag}_{\kappa_i}(t)\;] = 1  \;. \label{rela}
\end{equation}
Next, we build in a standard way the number operator
\begin{equation}
N_{\kappa_i}(t) = a^{\dag}_{\kappa_i}(t)a_{\kappa_i}(t) \;,
\label{number}
\end{equation}
which satisfies
\begin{equation}
[\;N_{\kappa_i}(t),a^{\dag}_{\kappa_i}(t)\;] =
a^{\dag}_{\kappa_i}(t)\;\;\;,\;\;\;
[\;N_{\kappa_i}(t),a_{\kappa_i}(t)\;] = -a_{\kappa_i}(t)   \;.
\label{rela11}
\end{equation}
Hence, one can identify  $a^{\dag}(t)$ and $a(t)$ objects with
creation and annihilation operators respectively; more preciously,
as we shall see for a moment, they create and annihilate the
time-dependent disc area quanta of space-times (\ref{spaces}).

It is easy to see that the number operator (\ref{number}) can be
rewritten in terms of coordinate variables  as follows
\begin{equation}
N_{\kappa_i}(t) = \frac{1}{2f_{\kappa_i}(t)} \left( [\;x_1\;]^2 +
[\;x_2\;]^2 -f_{\kappa_i}(t) \right) \;. \label{number11}
\end{equation}
Then, the  disc area  operator
\begin{equation}
S =  \pi \left( [\;x_1\;]^2 + [\;x_2\;]^2   \right) \;, \label{area}
\end{equation}
takes the form
\begin{equation}
S_{\kappa_i} (t) =  2 \pi f_{\kappa_i}(t) \left( N_{\kappa_i}(t)
+{1}/{2} \right) \;. \label{area11}
\end{equation}
 One can  find (see e.g. \cite{podrecznik}), that the operator  $S_{\kappa_i}(t)$ is
quantized with levels equally spaced by the  interval (quanta)
$s_{\kappa_i} (t) =2\pi f_{\kappa_i}(t)$, with the following
eigenvalues
\begin{equation}
S_{\kappa_i} (t)\varphi_{n,\kappa_i} = s_{n,\kappa_i} (t)
\varphi_{n,\kappa_i} = 2f_{\kappa_i}(t)\left(n + {1}/{2}
\right)\varphi_{n,\kappa_i} \;\;;\;\; n =0, 1, 2, \ldots
 \;,
\label{eigen}
\end{equation}
and the eigenstates given by
\begin{equation}
\varphi_{n,\kappa_i} (t) = \frac{1}{\sqrt{n!}} \left(
a^{\dag}_{\kappa_i}(t) \right)^n \varphi_{0,\kappa_i} \;\;;\;\;
a_{\kappa_i}(t)\varphi_{0,\kappa_i} =0
 \;.
\label{eigenstate}
\end{equation}
We see, therefore, that objects $a^{\dag}_{\kappa_i}(t)$ and
$a_{\kappa_i}(t)$ are (in fact) responsible for creation and
annihilation of area quanta  $s_{\kappa_i} (t)$ respectively.

Finally, let us study the basic properties of the above spectra.
First of all, it should be   noted that for fixed time parameter
$t$, all (twelve) time-dependent area quanta
\begin{eqnarray}
s_{\kappa_1}({t}) &=& s_{\pm,\kappa_1}\left(\frac{t}{\tau}\right) =
2\pi\kappa_1\,C_{\pm}^2 \left(\frac{t}{\tau}\right)\;, \label{sw2}\\
s_{\kappa_2}({t}) &=& s_{\pm,\kappa_2}\left(\frac{t}{\tau}\right)
=2\pi\kappa_2\tau\, C_{\pm} \left(\frac{t}{\tau}\right)S_{\pm}
\left(\frac{t}{\tau}\right) \;, \label{sw3}\\
s_{\kappa_3}({t}) &=& s_{\pm,\kappa_3}\left(\frac{t}{\tau}\right)
=2\pi \kappa_3\tau^2\, S_{\pm}^2 \left(\frac{t}{\tau}\right) \;,
\label{sw4}\\
s_{\kappa_4}({t}) &=&
 s_{\pm,\kappa_4}\left(\frac{t}{\tau}\right) = 8\pi\kappa_4
 \tau^4\left(C_{\pm}\left(\frac{t}{\tau}\right)
-1\right)^2 \;, \label{sw5}\\
s_{\kappa_5}({t}) &=& s_{\pm,\kappa_5}\left(\frac{t}{\tau}\right) =
\pm 2\pi \kappa_5\tau^2 \left(C_{\pm}\left(\frac{t}{\tau}\right)
-1\right)C_{\pm} \left(\frac{t}{\tau}\right)\;, \label{sw6}
\end{eqnarray}
and
\begin{eqnarray}
 s_{\kappa_6}({t}) &=&
s_{\pm,\kappa_6}\left(\frac{t}{\tau}\right) = \pm 2\pi
\kappa_6\tau^3 \left(C_{\pm}\left(\frac{t}{\tau}\right)
-1\right)S_{\pm} \left(\frac{t}{\tau}\right)\;, \label{sw7}
\end{eqnarray}
become the same as mentioned above "canonical" quanta $s_{\theta}
=2\pi \theta $. Such a situation appears for
\begin{eqnarray}
t_{\kappa_1} &=& t_{+/-,\kappa_1} = -\tau\, {\rm arccosh}/\arccos
\left(-\sqrt{\frac{\theta}{\kappa_1}}\right)\;, \label{time2}\\
t_{\kappa_2} &=& t_{+/-,\kappa_2} = \frac{\tau}{2}\, {\rm
arcsinh}/\arcsin \left({\frac{2\theta}{\tau\kappa_2}}\right)  \;,
\label{time3}\\
t_{\kappa_3} &=& t_{+/-,\kappa_3} = -\tau\, {\rm arcsinh}/\arcsin
\left(\sqrt{\frac{\theta}{\tau^2\kappa_3}}\right) \;,
\label{time4}\\
t_{\kappa_4} &=& t_{+/-,\kappa_4} = -\tau\, {\rm arccosh}/\arccos
\left(-\sqrt{\frac{\theta}{4\tau^4\kappa_4 }}+1\right) \;,
\label{time5}\\
t_{\kappa_5} &=& t_{+/-,\kappa_5} = -\tau\, {\rm arccosh}/\arccos
\left(\frac{1+/-\sqrt{1+/-\frac{4\theta}{\tau^2\kappa_5}}}{2}\right) \;,
\label{time6}
\end{eqnarray}
\begin{eqnarray}
t_{\kappa_6} &=& t_{+/-,\kappa_6}
=-\tau\, {\rm arccosh}/\arccos \left(\frac{1}{2} + \frac{1}{2}\sqrt{1-/+A_{+/-}+B_{+/-}}\; +\right.~~~~~\nonumber\\
&~&~~~~~~~~-\;\left.\frac{1}{2}
\sqrt{2+/-A_{+/-}-B_{+/-}-\frac{2}{\sqrt{1-/+A_{+/-}+B_{+/-}}}}\right)\;,\label{time7}
\end{eqnarray}
with
\begin{eqnarray}
A_{+/-}&=&\frac{2\times 2^{2/3}\left(\frac{\theta}{\kappa_6}
\right)^2}{\sqrt[3]{3}\sqrt[3]{-/+9\left(\frac{\theta}{\kappa_6}
\right)^2\tau^{12}-/+\sqrt{3}\sqrt{+/-16\left(\frac{\theta}{\kappa_6}
\right)^6\tau^{18}+ 27\left(\frac{\theta}{\kappa_6}
\right)^4\tau^{24}}}}\;, \nonumber\\
B_{+/-}&=& \frac{\sqrt[3]{2}\sqrt[3]{-/+9\left(\frac{\theta}{\kappa_6}
\right)^2\tau^{12}-/+\sqrt{3}\sqrt{+/-16\left(\frac{\theta}{\kappa_6}
\right)^6\tau^{18}+ 27\left(\frac{\theta}{\kappa_6}
\right)^4\tau^{24}}}}{3^{2/3}\tau^6}\;,
\nonumber
\end{eqnarray}
in the case of acceleration-enlarged Newton-Hooke Hopf structures
$\,{\mathcal U}_{\kappa_1}(\widehat{ NH}_{\pm})$, $\,{\mathcal
U}_{\kappa_2}(\widehat{ NH}_{\pm})$, $\,{\mathcal
U}_{\kappa_3}(\widehat{ NH}_{\pm})$, $\,{\mathcal
U}_{\kappa_4}(\widehat{ NH}_{\pm})$, $\,{\mathcal
U}_{\kappa_5}(\widehat{ NH}_{\pm})$ and $\,{\mathcal
U}_{\kappa_6}(\widehat{ NH}_{\pm})$ respectively.  Besides, one can
also observe that  first, third, fourth and fifth
acceleration-enlarged Newton-Hooke spectrum is invariant with
respect time reflection $t \to -t$, whereas such a symmetry is
broken in the case of second and sixth quanta $s_{\kappa_2}({t})$
and $s_{\kappa_6}({t})$. The last interesting property concerns the
"duality" of functions $s_{\kappa_1}({t})$, $s_{\kappa_2}({t})$ and
$s_{\kappa_3}({t})$ respectively, i.e. it should be noted that the
first of them passes into the last one (and vice-versa) after
substitution
\begin{equation}
C_{\pm} \left(\frac{t}{\tau}\right) \;\leftrightarrow \;\tau S_{\pm}
\left(\frac{t}{\tau}\right)\;, \label{duality}
\end{equation}
while the second quanta $s_{\kappa_2}({t})$ passes into
itself\footnote{In other words it is  "self-dual" with respect
"duality" transformation (\ref{duality}).}. Obviously, for
parameters $\kappa_i$ approaching infinity all deformations
disappear and the area spectra become classical.

In summary. In this short article we derive the spectra of disc area
operator defined on acceleration-enlarged Newton-Hooke space-times,
equipped with quantum spatial directions $x_i$ and classical time
$t$. We demonstrate that such obtained spectra are quantized with
level equally spaced by time-dependent area quanta, proportional to
the "value of noncommutativity", i.e.
to the function $f(t)$. Besides, it should be noted,  that the above 
quanta   are covariant with respect  the action of corresponding
twist deformed acceleration-enlarged Newton-Hooke Hopf symmetries,
respectively. Such a property obviously
 follows from the fact, that considered quantum space-times  are defined
as Hopf modules (representation spaces) of proper quantum groups.
The more physical applications for obtained results are under
investigation.

 \section*{Acknowledgments}
The author would like to thank J. Lukierski
for valuable discussions.\\
This paper has been financially supported by Polish Ministry of
Science and Higher Education grant NN202318534.


\begin{thebibliography}{99}
\bibitem{mech}A. Deriglazov, JHEP 0303, 021 (2003)
\bibitem{mechnext}S. Ghosh, Phys. Lett. B 648, 262
(2007)
\bibitem{qm}V.P. Nair, A.P. Polychronakos, Phys. Lett. B 505, 267 (2001)
\bibitem{qmnext}M. Chaichian, M.M. Sheikh-Jabbari, A. Tureanu, Phys.
Rev. Lett. 86, 2716 (2001)
\bibitem{field}M. Chaichian, P. Pre\v{s}najder and  A. Tureanu,
Phys. Rev. Lett. 94, 151602 (2005)
\bibitem{fieldnext}G. Fiore, J. Wess, Phys. Rev. D
75, 105022 (2007)
\bibitem{snyder}H.S. Snyder, Phys. Rev. 72, 68 (1947)
\bibitem{grav1}S. Doplicher, K. Fredenhagen, J.E. Roberts, Phys. Lett. B 331, 39
(1994)
\bibitem{string1}A. Connes, M.R. Douglas, A. Schwarz, JHEP 9802, 003
(1998)
\bibitem{1a}S. Coleman, S.L. Glashow, Phys. Rev. D 59, 116008
(1999)
\bibitem{1anext}
R.J. Protheore, H. Meyer, Phys. Lett. B 493, 1 (2000)
\bibitem{class1}S. Zakrzewski
; q-alg/9602001
\bibitem{class2}
Y. Brihaye, E. Kowalczyk, P. Maslanka
; math/0006167
\bibitem{oeckl}R. Oeckl, J. Math. Phys. 40, 3588 (1999)
\bibitem{chi}M. Chaichian, P.P. Kulish, K. Nashijima, A. Tureanu, Phys. Lett. B
604, 98 (2004)
\bibitem{dasz1}M. Daszkiewicz,
Mod. Phys. Lett. A 23, 505 (2008)
\bibitem{kappaP}J. Lukierski, A. Nowicki, H. Ruegg and V.N. Tolstoy, Phys. Lett.
B 264, 331 (1991)
\bibitem{kappaG}S. Giller, P. Kosinski, M. Majewski, P. Maslanka
and J. Kunz, Phys. Lett. B 286, 57 (1992)
\bibitem{lie1}
J. Lukierski and M. Woronowicz, Phys. Lett. B 633, 116 (2006)
\bibitem{qdef}O. Ogievetsky, W.B.  Schmidke, J. Wess, B. Zumino, Comm. Math. Phys.
150, 495 (1992)
\bibitem{paolo}
P. Aschieri, L. Castellani, A.M. Scarfone, Eur. Phys. J. C 7, 159
(1999)
\bibitem{nh}M. Daszkiewicz, Acta Phys. Pol. B 41, 1889 (2010)
\bibitem{nh1}M. Daszkiewicz, Mod. Phys. Lett. A 24, 1325 (2009)
\bibitem{bloch}C. Blohmann, J. Math. Phys. 44, 4736 (2003)
\bibitem{wess}J. Wess
; hep-th/0408080
\bibitem{fractal}D. Benedetti, Phys. Rev. Lett. 102, 111303 (2009)
\bibitem{dsr}G. Amelino-Camelia, Mod. Phys. Lett. A 17, 899
(2002)
\bibitem{area}J.M. Romero, J.A. Santiago, J.D. Vergara, Phys. Rev. D
68, 067503 (2003)
\bibitem{QG}C. Rovelli, \textit{"Quantum Gravity"}, UK: Univ. Pr. (2004)
\bibitem{spinfoam}J.C. Baez, J.D. Christensen, T.R. Halford, D.C.
Tsang, Class. Quantum Grav. 19, 4627 (2002)
\bibitem{spinfoam1}A. Alekseev, A.P. Polychronakos, M. Smedbck;
hep-th/0004036
\bibitem{snyderspec}J.M. Romero, A. Zamora, Phys. Lett. B 661, 11
(2008)
\bibitem{bekenstein}J.D. Bekenstein, Lett. Nuovo Cimento 11, 467 (1974)
\bibitem{podrecznik}J. Rzewuski, \textit{"Introduction to Quantum
Theory"}, Wroclaw Poli. Pr. (1992)

\end{thebibliography}
\end{document}